\documentclass[12pt,twoside,openany]{paper}
\usepackage{enumerate,amsmath,graphics,amssymb,graphicx,amscd,xypic,amscd,amsbsy,multirow,
float,booktabs,verbatim,pstricks,pst-plot,pst-node,color,xcolor,natbib}
\usepackage{hyperref}

\newcommand{\QED}{\hspace*{\fill}\rule{2.5mm}{2.5mm}}

\newcommand{\R}{\mathbb{R}}

\begin{document}
\begin{center}
\vspace{0.5cm} {\large \bf Utilizing wind in spatial covariance}\\
{Reza Hosseini}
\end{center}

\newcommand{\adr}{C://Dropbox//Public//Research//USC_research//exposure//figs//}

\begin{abstract}
This work develops a covariance function which allows for a stronger spatial correlation for pairs of points in the
direction of a vector such as wind and weaker for pairs which are perpendicular to it. It derives a simple covariance function by
stretching the space along the wind axes (upwind and across wind axes). It is shown that this covariance function
is anisotropy in the original space and the functions is explicitly calculated. 

\begin{comment}Then a simulation method is
proposed for air pollution transport (by wind) and the proposed covariance function is fitted to the simulations.
\end{comment}

\end{abstract}

\section{Introduction}

Many spatial processes show non-stationary behavior in the correlation function across space. Therefore various
non-stationary models are proposed in the literature to deal with this non-stationarity. A classical work in this
area is \cite{sampson-1992} which utilizes non-parametric {\it space transformations} of the space where the
process is transformed into another space where the process is almost isotropy. In the presence of wind, we
can expect a specific type of non-stationarity in which the correlation is larger between a pair of points along
the wind direction (angle=0) and lower between a pair of points 
perpendicular to the wind direction (with the same distance as before). We can also expect that this non-stationarity varies
smoothly as we vary the angle from 0 to $\pi/2$.

In this work using an intuitively appealing space transformation which satisfies the aforementioned expectation
we derive the exact form of the covariance function for a wind of given velocity and direction (Section
\ref{sect:covariance-derivation}). It turns out that the covariance function is an anisotropic covariance function (see \cite{banerjee-2003}) of a specific form which we calculate here.

\begin{comment}
In order to test if such a model is capturing the covariance function of an air pollution process in the presence
of wind, in Section \ref{sect:wind-transport-simulation}, we develop a method to simulate air pollution transport
using Brownian motion and point processes. The simulations is done for sources with fixed location or moving
location. Using these simulations we can calculate the empirical correlation function and check if it can be
well-approximated by the covariance functions we proposed in this work. Section
\ref{sect:fitting-covariance-derivation} uses hierarchical Bayesian spatial models (\cite{banerjee-2003}) for
estimating the parameters of the covariance function.
\end{comment}

\section{Covariance derivation}
\label{sect:covariance-derivation}
\begin{figure}[H]
\centering
\includegraphics[width=0.4\textwidth]{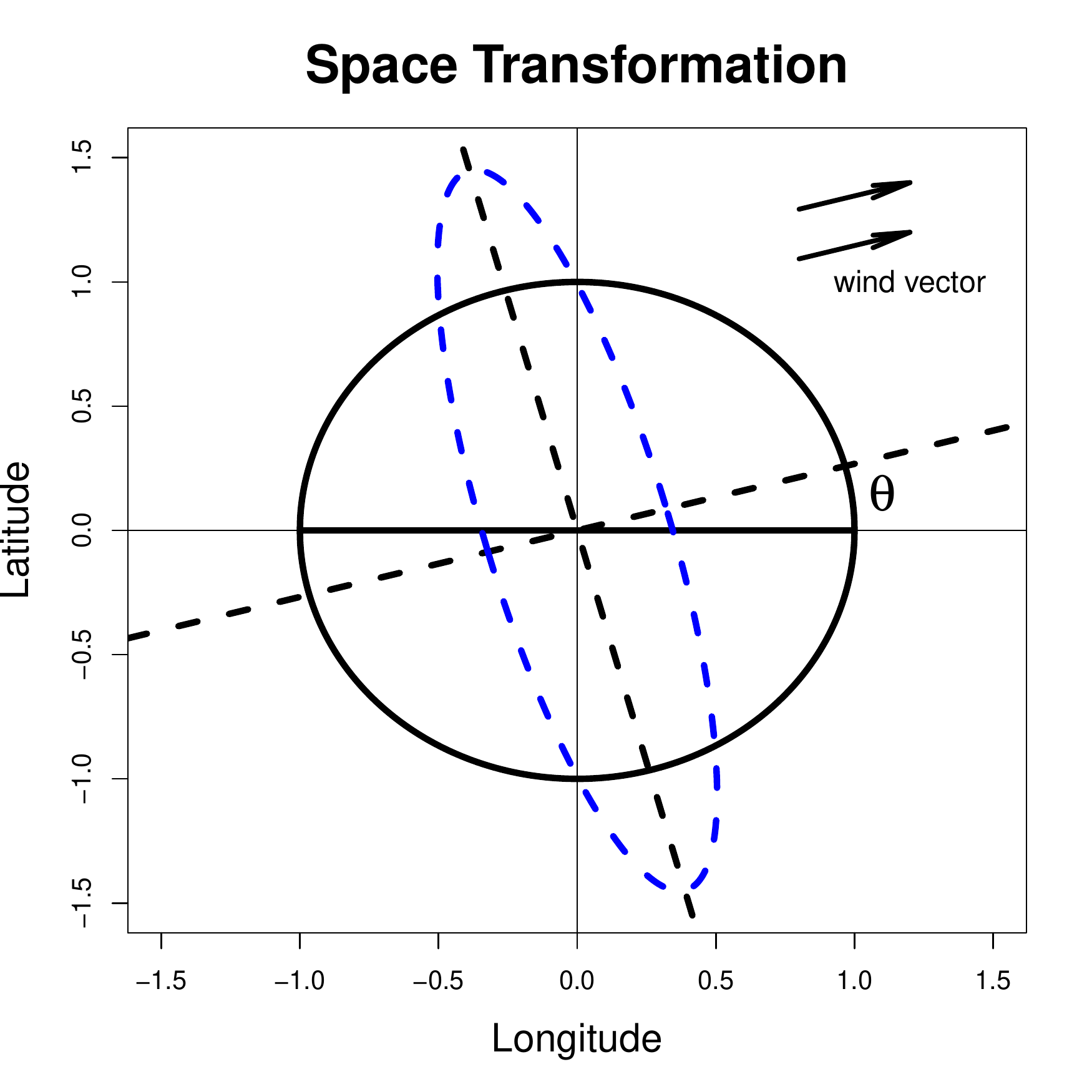}
 \caption{The space deformation for $a=3, b=2/3, \theta=\pi/12$. The original space is considered to be the circle depicted in bold and the deformed space is given with in dashed. The bold perpendicular lines depict the original axes and the dashed perpendicular lines depict the new axes.}
 \label{wind_cov_construct.pdf}
\end{figure}

 We denote the wind speed vector by ${\bf v}$ and assume it is constant across the domain we consider. Here we only consider a 2-dimensional vector but it is
possible to extend this method to 3 dimensions (and more). We denote the magnitude of the wind by $v$ and its angle with
the $x$ axis by $\theta_0$. It is useful to consider a new set of coordinates with the same origin and x-axis with the same direction as wind and y-axis perpendicular to it. We denote the coordinates of a point $(x,y)$ in the original plane
by $(x^*,y^*)$ and in the new coordinates which we call the {\it wind coordinates}. The idea of the method is depicted in Figure \ref{wind_cov_construct.pdf} where the original space is considered to be the circle depicted in bold and the deformed space is given with in dashed. In the new space the transformed pair of points for which the connecting line segment is parallel to the wind direction are closer than the original space. We can find the transformation to achieve this result by stretching and rotation of the space in three steps: (1) rotate the space counterclockwise with the angle $\theta_0$ to find the coordinates of the points in the wind coordinates; (2) stretch the space along the wind axes; (3) rotate back the results to the get the value of the transformation in the original space.

Let us denote the counter-clockwise rotation matrix of angle $\theta_0$ by $R(\theta_0)$ which is given by
\[R(\theta_0)=\left(\begin{array}{cc}\cos(\theta_0) & -\sin(\theta_0)\\
\sin(\theta_0) & \cos(\theta_0)\\  \end{array}\right)\]

Also let $a,b\geq 0$ denote the magnitude of the {\it stretch} parallel to the
wind direction and perpendicular to it respectively. Denote its matrix by $S(a,b)$ which is given by
\[S(a,b)=\left(\begin{array}{cc}1/a & 0\\
0 & 1/b\\  \end{array}\right).\] 
Suppose $(x,y)$ is given in the original space with $(x^*,y^*)$ denoting the same point in the wind coordinates:
\begin{equation}
(x^*,y^*)^t = R(\theta_0) (x,y)^t,
\label{eqn-trans}
\end{equation}
where $t$ denotes the matrix transpose operation.
 We assume that the point $(x,y)$
is stretched along the axes of the wind coordinates:
\[\left(\begin{array}{c}x^* \\ y^* \\  \end{array}\right) \mapsto \left(\begin{array}{c}x^*/a \\ y^*/b \\  \end{array}\right)=S(a,b)\left(\begin{array}{c}x^* \\ y^* \\  \end{array}\right) \]
We are interested to find the transformation formula in the original space. In order to do so, we find the coordinates of
$(x^*,y^*)$ in the original space. To that end, we can apply the rotation  $R(-\theta_0)=R(\theta_0)^t$ (where $t$ denotes matrix transpose). Hence by Equation \ref{eqn-trans}, the above mapping can be
written in the original space as
\[R(\theta_0)^t \left(\begin{array}{c}x \\ y \\  \end{array}\right) \mapsto  S(a,b)R(\theta_0)^t \left(\begin{array}{c}x \\ y \\  \end{array}\right).\]
By multiplying $(R(\theta_0)^t)^{-1}=R(\theta_0)$ to the two sides, we can write the above mapping in the original space as 
\[\left(\begin{array}{c}x \\ y \\  \end{array}\right) \mapsto  R(\theta_0)S(a,b)R(\theta_0)^t \left(\begin{array}{c}x \\ y \\  \end{array}\right).\]

We call $\gamma=a/b$ the {\it relative stretch} parameter. Then we can write
\[A=(1/b)R(\theta_0)S(\gamma,1)R(\theta_0)^t,\]
we denote $R(\theta_0)S(\gamma,1)R(\theta_0)^t$ by $A[\gamma]$ which is a symmetric matrix.

The hope is that the covariance in the new space has a simple form so that after applying this
transformation, we can model the covariance appropriately using a simple model. We assume that in the new
space the covariance is isotropic, i.e. it is only a function of the Euclidean distance. The Euclidean distance between two points is given by
\[{\bf s}=\left(\begin{array}{c}x \\ y \\  \end{array}\right),\;\;{\bf s'}=\left(\begin{array}{c}x' \\ y' \\  \end{array}\right)\]
which in the original space is equal to 
\[d_E(\bf{s},\bf{s'})=  {\bf h}^t {\bf h},\]
where ${\bf h}={\bf s} - {\bf s'}.$ However as indicated we would like to use the Euclidean
distance in the new
space as the distance used in isotropic covariance models:
\[d({\bf s},{\bf s'})=d_E(A{\bf s},A{\bf s'}) = (A{\bf h})^t (A{\bf h})=\sqrt{{\bf h}^t A^tA {\bf h}}=(1/b^2)\sqrt{{\bf h}^t A[\gamma]^tA[\gamma]} {\bf h}.\]
Note that 
\[A[\gamma]^tA[\gamma] = R(\theta_0) S(\gamma,1) R(\theta_0)^tR(\theta_0) S(\gamma,1) R(\theta_0)^t=
A[\gamma^2]\]

\begin{figure}[H]
\centering
\includegraphics[width=0.4\textwidth]{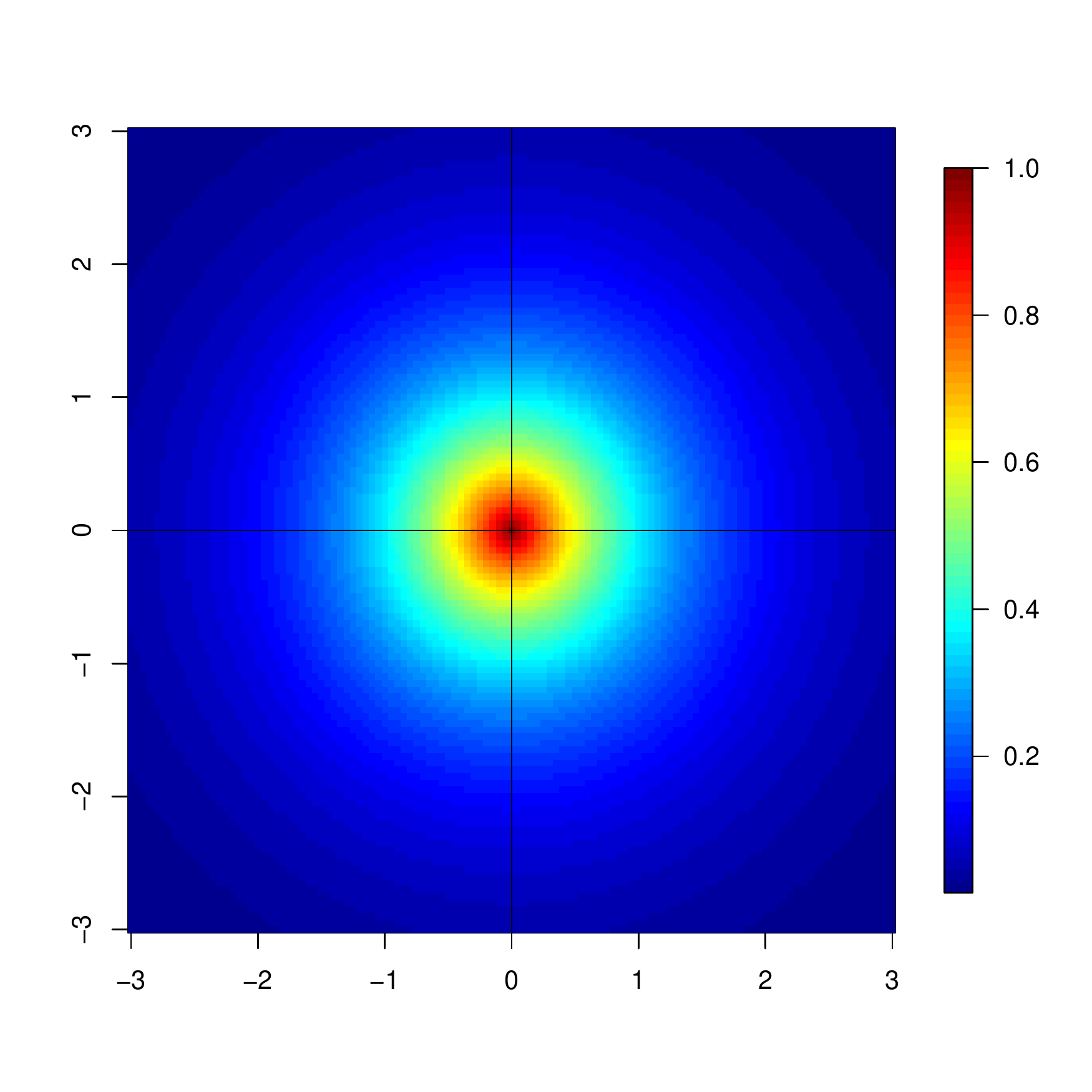}\includegraphics[width=0.4\textwidth]{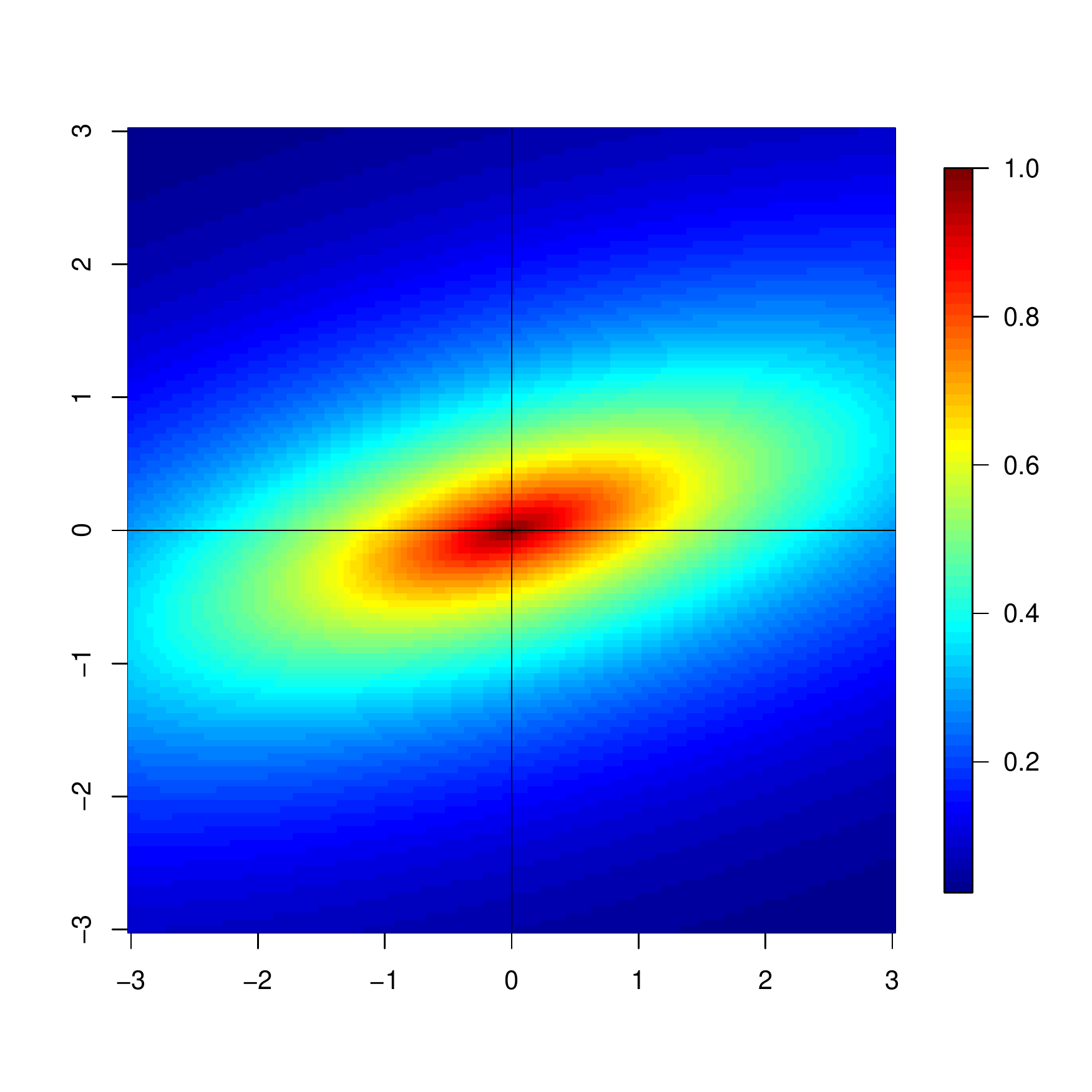}
 \caption{The exponential covariance function transformation for a=3,b=1, $\theta=\pi/12$.
 The left panel is the original covariance function and the right panel is the covariance
 function of the deformed space.}
 \label{wind_cov_example.pdf}
\end{figure}

As an example consider a gaussian covariance function: 
\[C({\bf h}) =\sigma^2 \exp(-{\bf h}^t{\bf h}/\phi),\] the new covariance function is given by 
\[C(s,s') =\sigma^2 \exp(-({\bf h}^tA[\gamma^2]{\bf h})/(\phi b^2)),\]
which implies that the parameter $b$ is not identifiable and can be absorbed into the range parameter $\phi$ by the change of variables $(\phi b^2) \mapsto \phi $ to arrive at the anisotropy covariance function:
\[C(s,s') =\sigma^2 \exp(-{\bf h}^tA[\gamma^2]{\bf h}/\phi).\]
Figure \ref{wind_cov_example.pdf} depicts the application of this method to the exponential covariance function 
$C(s,s') = \exp(-({\bf h}^t{\bf h})^{1/2}).$

\section{Conclusion}

We obtained a simple covariance function which achieves higher spatial correlation for pairs of points in the
direction of wind and lower for pairs which are perpendicular and give the closed-form formula. It turns out that the covariance function is anisotropy and except for the wind angle depends on only one other parameter which we called the relative stretch parameter, $\gamma$. The relative stretch can be considered to be a function of the wind speed for modeling, for example by letting 
$\gamma = \exp(v \gamma')$. In that case $\gamma'=0$ correspond to no stretch case which is also the case when the wind speed is zero ($v=0$) as desired. $\gamma'>0$ corresponds to a stretch perpendicular to the wind vector and $\gamma'<0$ corresponds to a stretch parallel to the wind vector. The same formulation can be used for transforming a kernel for averaging a predictor such as a traffic variable which is a proxy for pollution source emission. In that context we can transform an isotropic kernel (with circle contours) which is utilized for averaging the source effect around a given spatial point $s$ to a kernel which is directed parallel to the wind (with ellipse contours) at that point. The resulting kernel in that case will have exactly the same form as we discussed here.

{\noindent \bf Acknowledgements}: 
The author gratefully acknowledges useful discussions with Drs Duncan Thomas, Kiros Berhane, and Meredith Franklin from University of Southern California. This work was partially supported by National Institute of Environmental Health Sciences (5P30ES007048, 5P01ES011627, 5P01ES009581); United States Environmental Protection Agency (R826708-01, RD831861-01); National Heart Lung and Blood Institute (5R01HL061768, 5R01HL076647); California Air Resources Board contract (94-331); and the Hastings Foundation.

\clearpage

\newpage

\end{document}